\documentstyle[aps,floats,psfig,epsf,twocolumn]{revtex}
\def\gz{\ifmmode{Z\hskip -4.8pt Z}
    \else{\hbox{$Z\hskip -4.8pt Z$}}\fi}

\newcommand{\be}{\begin{equation}}
\newcommand{\ee}{\end{equation}}
\newcommand{\bea}{\begin{eqnarray}}
\newcommand{\eea}{\end{eqnarray}}

\begin{document}
\tighten
\draft
\title{Theoretical analysis of magnetic Raman scattering in
La$\mbox{\boldmath$_2$}$CuO$\mbox{\boldmath$_4$}$:
two-magnon intensity with the inclusion of ring exchange}
\author{A.~A. Katanin$^{a,b}$ and A.~P. Kampf $^a$}
\address{\sl
$^a$ Institut f\"ur Physik, Theoretische Physik III,
Elektronische Korrelationen und Magnetismus,\\
Universit\"at Augsburg, 86135 Augsburg, Germany\\
$^b$ Institute of Metal Physics, 620219 Ekaterinburg, Russia}
%\date{\today}
\address{~
\parbox{14cm}{\rm
\medskip
We evaluate the Raman light scattering intensity for the square lattice
Heisenberg antiferromagnet with plaquette ring exchange $J_{\Box}$. With
the exchange couplings as fixed before from an accurate fit to the spin
wave dispersion in La$_2$CuO$_4$, leading in particular to $J_{\Box}=0.24J$,
we demonstrate in a parameter free calculation that the inclusion of the
plaquette exchange contribution to the dispersion and the magnon-magnon
interaction vertex gives a peak position in $B_{1g}$ scattering geometry
$E_{max} = 2.71J$ which is in excellent agreement with the experimental data.
Yet, the intrinsic width and the lineshape of the two-magnon remain beyond a
descriptions in terms of a spin-only Hamiltonian.
}}

\maketitle

The magnetic properties of La$_2$CuO$_4$ have been the subject of many
detailed investigations over the last decade. Understanding this undoped
parent compound of high temperature superconducting cuprates is a
precondition for the many theories which describe metallic cuprates by
doping carriers into a layered antiferromagnet. The conventional starting
point for undoped cuprates is the two-dimensional (2D) spin-1/2 Heisenberg
model with nearest-neighbor (nn) exchange interaction $J$ \cite{Manousakis}.
Despite the substantial progress on the theory of the 2D Heisenberg
antiferromagnet \cite{Chakravarty}, some of the experimental facts for La$_2$%
CuO$_4$ have clearly demonstrated that a complete description of the
magnetic excitations requires additional physics not contained in the 2D
Heisenberg model with $J$ only. Examples include the asymmetric lineshape of
the two-magnon Raman intensity \cite{Lyons} or the infrared optical
absorption \cite{Perkins}.

The importance of an additional ring (plaquette) exchange coupling for La$_2$%
CuO$_4$ recently found direct experimental support from the observed
dispersion of the spin-waves along the magnetic Brillouin zone boundary \cite
{Aeppli}. A fit of the experimental results \cite{Aeppli} using the
theoretical spin-wave dispersion, which consistently includes quantum
renormalization effects \cite{Our}, has provided accurate values for the nn
exchange integral $J=1720K$ and the ring-exchange coupling $\,J_{\Box
}=0.24J $. The new estimate for $J$ corrects previous values which have been
used in past years and which were consistently 10 to 15\% lower. Also the
value for $J_{\Box}$ must be considered surprisingly large, but the spin
stiffness and the N\'eel temperature calculated with the new parameter set
obtained in Ref. \cite{Our} were in excellent agreement with the
experimental data and thereby confirmed the deduced exchange coupling
parameters. The value of the ring exchange $J_{\Box}$ also agreed with the
strong-coupling expansion studies of the three-band Hubbard model in the
parameter range relevant for CuO$_2$ planes \cite{Band3}.

The Heisenberg model with $J$ only was previously found insufficient to
describe the experimentally observed asymmetry and width of the Raman
spectra of undoped cuprate compounds \cite{Lyons}. The Raman spectrum of the
Heisenberg model was investigated within the ladder approximation for the
magnon-magnon scattering vertex many years ago \cite{Dav,Canali} and gave an
almost symmetric narrow peak located at $E_{\max }=3.3J$. It was shown \cite
{Canali} that magnon self-energy and vertex corrections to the ladder
approximation as well as 4-magnon scattering contributions are small and
negligible. Numerically, the problem was investigated by series expansions
around the Ising limit \cite{Singh1}, exact diagonalizations on small
clusters \cite{Dagotto}, and QMC calculations \cite{Sandvik}. The results of
these studies led to a peak position and a two-magnon lineshape which were
very close to the spin-wave results. Furthermore, the value of the exchange
integral $J=1440$K for La$_2$CuO$_4$ as extracted from the position of the
two-magnon Raman peak appears too small in comparison to early neutron
scattering results for the spin-wave spectrum ($J=1650$K), see Ref. \cite
{SSWT} and references therein.

The asymmetric lineshape of the two-magnon Raman intensity \cite{Lyons} has
led to proposals that spin-phonon interactions \cite{Knoll,Nori,FreitasSingh}%
, resonant phenomena \cite{ChubukovFrenkel,Schonfeld}, purely fermionic
contributions \cite{Ho}, or cyclic ring exchange \cite
{Honda,Eroles,Lorenzana} need to be included beyond the nn Heisenberg model.
A possible importance of ring exchange for Raman scattering was conjectured
also from numerical calculations \cite{Roger89} which showed that a finite $%
J_{\Box}$ gives rise to additional high energy contributions in the Raman
intensity -- yet with little spectral weight.

The new, accurate estimates for the exchange couplings therefore demand a
theoretical reanalysis of two-magnon scattering in La$_2$CuO$_4$, which we
investigate in the present paper with an emphasis on the effects of ring
exchange. The question that we address is to what extent the new values for
the exchange parameters -- including the finite $J_{\Box}$ -- consistently
describe the two-magnon intensity in La$_2$CuO$_4$ and whether the
Heisenberg model with ring exchange alone is sufficient to explain the
experimentally observed lineshape of the Raman intensity. To answer this
question we calculate the magnon-magnon interaction vertex in the presence
of the ring exchange term and then solve the ladder equation for the Raman
scattering vertex.

We start from the Heisenberg model on a square lattice with ring-exchange
coupling \cite{Roger89,Takahashi77,MacDonald90}
\begin{eqnarray}
H &=&J\sum_{i,\delta }{\bf S}_i\cdot {\bf S}_{i+\delta }+J^{\prime
}\sum_{i,\delta ^{\prime }}{\bf S}_i\cdot {\bf S}_{i+\delta ^{\prime
}}+J^{\prime \prime }\sum_{i,\delta ^{\prime \prime }}{\bf S}_i\cdot {\bf S}%
_{i+\delta ^{\prime \prime }}  \nonumber \\
&&\ \ \ \ \ \ \ +J_{\Box }\sum_{\langle ijkl\rangle }\big[ ({\bf S}_i\cdot
{\bf S}_j)({\bf S}_k\cdot {\bf S}_l)  \nonumber \\
&&\ \ \ \ \ \ \ +({\bf S}_i\cdot {\bf S}_l)({\bf S}_k\cdot {\bf S}_j)-({\bf S%
}_i\cdot {\bf S}_k)({\bf S}_j\cdot {\bf S}_l)\big]  \label{Hamil}
\end{eqnarray}
where $J,\,J^{\prime }$, and $J^{\prime \prime }$ are the first ($\delta $),
second ($\delta ^{\prime }$) and third ($\delta ^{\prime \prime }$) nearest
neighbor in-plane exchanges. We use the Dyson-Maleev representation for the
spin operators
\begin{equation}
\left.
\begin{array}{l}
S_i^{+}=\sqrt{2S}a_i\,,\;S_i^z=S-a_i^{\dagger }a_i \\
S_i^{-}=\sqrt{2S}(a_i^{\dagger }-\frac 1{2S}a_i^{\dagger }a_i^{\dagger }a_i)
\end{array}
\right\} \,i\in A\,\,,  \label{BKJa}
\end{equation}
\begin{equation}
\left.
\begin{array}{l}
S_j^{+}=\sqrt{2S}b_j^{\dagger }\,,\;S_j^z=-S+b_j^{\dagger }b_j \\
S_i^{-}=\sqrt{2S}(b_i-\frac 1{2S}b_i^{\dagger }b_ib_i)
\end{array}
\right\} \,j\in B\,\,  \label{BKJb}
\end{equation}
where $A$ and $B$ denote the two sublattices of the antiferromagnet; $%
a_i^{\dagger },a_i,$ and $b_j^{\dagger },b_j$ are Bose operators and $S=1/2$%
. With the Bogoliubov transformation
\begin{eqnarray*}
a_{{\bf k}} &=&u_{{\bf k}}\alpha _{{\bf k}}+v_{{\bf k}}\beta _{{\bf k}%
}^{\dagger } \\
b_{{\bf k}} &=&u_{{\bf k}}\beta _{{\bf k}}+v_{{\bf k}}\alpha _{{\bf k}%
}^{\dagger }
\end{eqnarray*}
and by introducing the ``coherence factors''
\begin{eqnarray*}
u_{{\bf k}} &=&[(A_{{\bf k}}+E_{{\bf k}})/(2E_{{\bf k}})]^{1/2} \\
v_{{\bf k}} &=&-[(A_{{\bf k}}-E_{{\bf k}})/(2E_{{\bf k}})]^{1/2}
\end{eqnarray*}
we diagonalize the quadratic part of the Hamiltonian and obtain the
spin-wave spectrum (for details see \cite{Our})
\begin{eqnarray}
E_{{\bf k}} &=&\sqrt{A_{{\bf k}}^2-B_{{\bf k}}^2},  \label{Ek} \\
A_{{\bf k}} &=&4S[J\gamma -J^{\prime }\gamma ^{\prime }(1-\nu _{{\bf k}%
}^{\delta ^{\prime }})-J_{\Box }S^2(\gamma _0^{\Box }+\gamma _{\delta
^{\prime }}^{\Box }\nu _{{\bf k}}^{\delta ^{\prime }})]  \nonumber \\
&&\ \ \ \ \ \ \ \ \ \ \ \ -4J^{\prime \prime }S\gamma ^{\prime \prime
}(1-\nu _{{\bf k}}^{\delta ^{\prime \prime }})\,, \\
B_{{\bf k}} &=&4S(J\gamma -J_{\Box }S^2\gamma _\delta ^{\Box })\nu _{{\bf k}%
}^\delta \,,
\end{eqnarray}
with the momentum dependent coefficients
\begin{eqnarray}
\nu _{{\bf k}}^\delta  &=&(\cos k_x+\cos k_y)/2;\;\nu _{{\bf k}}^{\delta
^{\prime }}=\cos k_x\cos k_y;\;  \nonumber \\
\nu _{{\bf k}}^{\delta ^{\prime \prime }} &=&(\cos 2k_x+\cos 2k_y)/2\,.
\end{eqnarray}
Note that the quantum renormalization factors $\{\gamma _\delta \}$ in (\ref
{Ek}) take into account the renormalization of the magnon spectrum due to
quartic terms\cite{Our}. The resulting Hamiltonian reads
\begin{eqnarray}
\  &&H=\sum_{{\bf k}}E_{{\bf k}}(\alpha _{{\bf k}}^{\dagger }\alpha _{{\bf k}%
}+\beta _{{\bf k}}^{\dagger }\beta _{{\bf k}})  \label{H24} \\
\  &+&\sum V_{{\bf k}_1{\bf k}_2;{\bf k}_3{\bf k}_4}^{ijmn}:R_{{\bf k}%
_1,i}^{\dagger }R_{{\bf k}_2,j}^{\dagger }R_{{\bf k}_3,m}R_{{\bf k}%
_4,n}:\delta _{{\bf k}_1+{\bf k}_2,{\bf k}_3+{\bf k}_4}  \nonumber
\end{eqnarray}
where we used the vector notation
\begin{equation}
R_{{\bf k}}=(\alpha _{{\bf k}},\beta _{-{\bf k}}^{\dagger })
\end{equation}
and $i,j,m,n=1,2.$ The quartic part in Eq. (\ref{H24}) is normal ordered,
since all ``Hartree-Fock'' renormalizations are already absorbed in the
quadratic part of the Hamiltonian, cf. Ref. \cite{Rastelli}.

We use the effective Loudon-Fleury \cite{LF} Hamiltonian in $B_{1g}$
geometry for the coupling of the incoming and outgoing photons to the
localized nn spins which are involved in the two-spin flip Raman process
\begin{equation}
H_R=\Lambda \sum_j{\bf S}_j\cdot ({\bf S}_{j+\delta _y}-{\bf S}_{j+\delta
_x})\,;  \label{HR}
\end{equation}
$\delta _{x(y)}$ are unit vectors in the directions $x,y$ respectively, and $%
\Lambda $ is a coupling constant which includes the electric field vectors
of the two photons. For the calculation of the Raman light scattering
intensity in the ladder approximation for repeated magnon-magnon scattering
processes only the vertex $V_{{\bf kk}^{\prime };{\bf k}^{\prime }{\bf k}%
}^{(4)}= V_{{\bf kk}^{\prime};{\bf k}^{\prime}{\bf k}}^{1,2,1,2}$ is needed
\cite{Dav,Canali}. The complicated vertex function $V_{{\bf k}_1{\bf k}2;%
{\bf k}_3{\bf k}_4}^{1,2,1,2}$ was calculated with the help of computer
algebra. After Bogoliubov transformation the ring-exchange term in Eq. (1)
gives 1057 (!) contributions with different combinations of $\alpha $ and $%
\beta $ operators. The general result for the vertex $V_{{\bf k}_1{\bf k}_2;%
{\bf k}_2{\bf k}_4}^{(4)}$ is rather involved, however, for equal momenta it
simplifies to
\begin{equation}
V_{{\bf kk}^{\prime };{\bf k}^{\prime }{\bf k}}^{(4)}={\frac{B_{{\bf k-k}%
^{\prime }}(A_{{\bf k}}A_{{\bf k}^{\prime }}+E_{{\bf k}}E_{{\bf k}^{\prime
}})-A_{{\bf k-k}^{\prime }}B_{{\bf k}}B_{{\bf k}^{\prime }}-R_{\Box }}{2E_{%
{\bf k}}E_{{\bf k}^{\prime }}}}  \label{V4}
\end{equation}
where the effects of the ring-exchange coupling are included in the magnon
spectrum (\ref{Ek}) and in the additional vertex contribution
\begin{eqnarray}
R_{\Box } &=&8J_{\Box }\left[ B_{{\bf k}}B_{{\bf k}^{\prime }}(4+\nu _{{\bf k%
}}^{\delta ^{\prime }}+\nu _{{\bf k}^{\prime }}^{\delta ^{\prime }}+\nu _{%
{\bf k-k}^{\prime }}^{\delta ^{\prime }}) + \right. \\
&-&2A_{{\bf k}^{\prime }}B_{{\bf k}}(\nu _{{\bf k}^{\prime }}^\delta +\nu _{%
{\bf k,k-k}^{\prime }}^{\Box })-2A_{{\bf k}}B_{{\bf k}^{\prime }}(\nu _{{\bf %
k}}^\delta +\nu _{{\bf k-k}^{\prime }{\bf ,k}^{\prime }}^{\Box })  \nonumber
\\
&+&2A_{{\bf k}}A_{{\bf k}^{\prime }}(\nu _{{\bf k-k}^{\prime }}^\delta +\nu
_{{\bf k,k}^{\prime }}^{\Box })+2E_{{\bf k}}E_{{\bf k}^{\prime }}(-\nu _{%
{\bf k-k}^{\prime }}^\delta +\nu _{{\bf k,k}^{\prime }}^{\Box })\big]
\nonumber
\end{eqnarray}
and $\nu_{{\bf k,k}^{\prime}}^{\Box}=(\cos k_x\cos k_y^{\prime}+\cos k_y
\cos k_x^{\prime})/2$.

The first two terms in the numerator of Eq. (\ref{V4}) coincide with the
corresponding result for the 2D nn Heisenberg antiferromagnet \cite
{Dav,Canali} but with renormalized coefficients $A_{{\bf k}},$ $B_{{\bf k}}$
, and the renormalized magnon dispersion $E_{{\bf k}}$ in (\ref{Ek}). The
time-ordered response function of two-magnon Raman scattering is given by
\[
G(\omega )=\int\limits_{-\infty }^\infty dte^{i\omega t}\langle
T[H_R(t)H_R(0)]\rangle
\]
where $H_R(t)$ is the Heisenberg representation of $H_R,$ Eq. (\ref{HR}).
The Raman light scattering intensity is then determined by $I(\omega
)=-(1/\pi )\,{\rm Im}\,G(\omega )$. In the ladder approximation we obtain
the equations \cite{Dav,Canali}:
\begin{eqnarray}
G(\omega )=i\Lambda ^2\int \frac{d\omega ^{\prime }}{2\pi }\sum_{{\bf k}}%
\frac{A_{{\bf k}}f_{{\bf k}}}{E_{{\bf k}}} &&G_\alpha ^{(0)}({\bf k},\omega
+\omega ^{\prime })\cdot  \\
&&\ \ \cdot G_\beta ^{(0)}({\bf k},\omega ^{\prime })\Gamma _{{\bf k}%
}(\omega )\,,  \nonumber
\end{eqnarray}
\begin{eqnarray}
\Gamma _{{\bf k}}(\omega )=\frac{f_{{\bf k}}A_{{\bf k}}}{E_{{\bf k}}}-i\int
\frac{d\omega _1}{2\pi }\sum_{{\bf k}_1} &&V_{{\bf kk}_1;{\bf k}_1{\bf k}%
}^{(4)}G_\alpha ^{(0)}({\bf k}_1,\omega +\omega _1)\cdot   \nonumber \\
&&\ \ \cdot G_\beta ^{(0)}({\bf k}_1,\omega _1)\Gamma _{{\bf k}_1}(\omega )
\end{eqnarray}
where $G_{\alpha ,\beta }^{(0)}({\bf k},\omega )=-1/(E_{{\bf k}}\mp \omega
-i0^{+})$ and $f_{{\bf k}}=\cos k_x-\cos k_y.$ The result reads
\begin{equation}
G(\omega )={\displaystyle{\frac{\Lambda ^2L_2(\omega )}{D(\omega )}}}\left\{
1+{\frac J2}\left[ L_0(\omega )-\displaystyle{\frac{L_1^2(\omega )}{%
L_2(\omega )}}\right] \right\} \,,
\end{equation}
\begin{eqnarray}
D(\omega ) &=&1+(J-2J_{\Box }S^2){\frac{L_2(\omega )}2}+(J+6J_{\Box }S^2){%
\frac{L_0(\omega )}2}  \nonumber \\
&&\ +(J-2J_{\Box }S^2)(J+6J_{\Box }S^2){\frac{L_0(\omega )L_2(\omega
)-L_1^2(\omega )}4}  \nonumber
\end{eqnarray}
where
\begin{eqnarray}
L_m(\omega ) &=&i\int \frac{d\omega ^{\prime }}{2\pi }\sum_{{\bf k}}\frac{f_{%
{\bf k}}^2A_{{\bf k}}^m}{E_{{\bf k}}^m}G_\alpha ^{(0)}({\bf k},\omega
+\omega ^{\prime })G_\beta ^{(0)}({\bf k},\omega ^{\prime })  \nonumber \\
\  &=&\sum_{{\bf k}}\frac{f_{{\bf k}}^2A_{{\bf k}}^m}{E_{{\bf k}}^m}\frac 1{%
\omega -2E_k}\,.  \label{Lm}
\end{eqnarray}
For $J_{\Box }=0$ the above formulas reduce to the known results of Refs.%
\cite{Dav,Canali}.

For the numerical calculations we use the parameter set determined in Ref.
\cite{Our} from the fit to the spin wave dispersion
\begin{equation}
J=151.9\,\text{meV},\,\,J^{\prime }=J^{\prime \prime }=0.025J,\,\,J_{\Box
}=0.24J.  \label{Js}
\end{equation}
For these exchange couplings the renormalization parameters follow as \cite
{Our}
\begin{eqnarray}
\gamma &=&1.158;\;\gamma ^{\prime }=0.909;\;\gamma ^{\prime \prime }=0.852
\label{gn} \\
\gamma _0^{\Box } &=&2.220;\;\gamma _\delta ^{\Box }=1.971;\;\gamma _{\delta
^{\prime }}^{\Box }=1.721  \nonumber
\end{eqnarray}
Nevertheless we first explore the $J_{\Box}$ dependence of the two-magnon
Raman intensity as shown in Fig. 1. On varying $J_{\Box}$ the
renormalization parameters (\ref{gn}) necessarily have to be recalculated
each time. For $J^{\prime }=J^{\prime \prime }=\,J_{\Box }=0$ the result in
Fig. 1 coincides with that obtained in Refs. \cite{Dav,Canali}. In a first
step we turn on $J^{\prime}$ and $J^{\prime\prime}$ and observe a shift of
the two-magnon peak position to lower frequencies and a further downward
shift when also the the ring-exchange coupling is added. The magnitude of
the peak shift due to the ring exchange is quite strong. In addition, due to
the finite $J_{\Box}$ the spectrum near the upper edge of the two-magnon
continuum develops a ``foot'' structure. For comparison, we plot also the
Raman scattering intensity for a twice larger ring exchange $J_{\Box }=0.48J$
where the above effects are even more pronounced. The position of the peak
estimated with $J_{\Box}=0.24J $ is $E_{\max }=2.71J$ and with the absolute
values in (\ref{Js}) is therefore excellent agreement with the experimental
result \cite{Lyons} $E_{\max }=2.65J=3150{\rm cm}^{-1}.$ If we turn off the
vertex contribution $R_{\Box }$ in Eq. (\ref{V4}) and therefore keep only
the influence of the ring exchange in the spin-wave dispersion, we obtain $%
E_{\max }=2.86J$ which is still much smaller than the peak position without
ring exchange ($E_{\max }=3.15J$). Therefore, the modification of the
spectrum due to a finite $J_{\Box}$ is the major origin for the peak shift
to lower energies.

\begin{figure}[t!]
%%\vspace{5mm}
\psfig{file=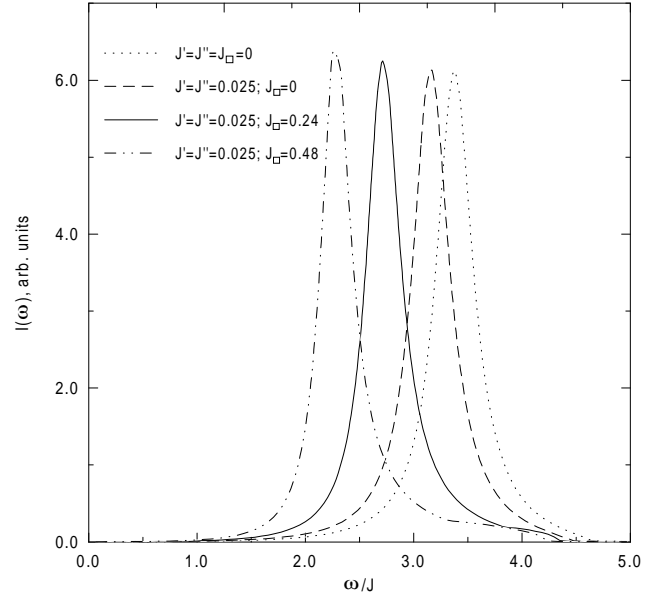,width=90mm,silent=}
%%\centerline{\psfig{file=Fig1.ps,width=90mm,silent=}}
\vspace{2mm}
\caption{Evolution of the two-magnon Raman intensity with increasing
ring-exchange couplings; different parameter sets are indicated in the
figure.}
\label{fig:JBoxdependence}
\end{figure}

In Fig. 2 we compare the Raman lineshape, obtained for the parameter set (%
\ref{Js}) with the experimental data on La$_2$CuO$_4$ from Ref. \cite{Lyons}%
. Although the precise peak position is in excellent agreement with the
data, its width is only slightly influenced by ring-exchange coupling. The
foot structure near the upper edge of the spectrum does result from ring
exchange and therefore indeed leads to an asymmetry of the lineshape, its
weight, however, is too small to account for the overall linewidth and
lineshape. Therefore, other processes must contribute which are not
described by the Heisenberg model with ring exchange alone. It was proposed,
that the damping due to spin-phonon coupling \cite{Knoll,Nori,FreitasSingh}
is responsible for the peak width which in the current calculation is
underestimated by roughly a factor 2. In order to incorporate extrinsic
sources for damping beyond the extended Heisenberg Hamiltonian Eq. (1), we
have introduced a small damping $\Gamma$ into the denominators of $%
L_m(\omega ),$ Eq. (\ref{Lm}). As a reasonable choice $\Gamma$ is assumed
constant over the entire Brillouin zone, because it is predominantly the
zone boundary magnons which determine the shape of the two-magnon peak. We
plot the result of this calculation for $\Gamma =0.13J$ in Fig. 2. One can
see that while the width of the experimentally observed spectra can be
fitted in such a way, the asymmetry of the spectrum is not accounted for.
Since the Raman spectra do depend on the incoming light frequency, the
two-magnon scattering is identified as a resonant process and therefore an
additional source for the strong asymmetry of the lineshape is expected to
arise from the coupling to the charge degrees of freedom \cite
{ChubukovFrenkel,Schonfeld}.

In summary, we have investigated the effect of ring exchange on two-magnon
Raman scattering in the 2D Heisenberg antiferromagnet. Using the previously
fixed exchange coupling parameter set we find in a fit-para\-meter free
calculation an excellent agreement with the position of the two-magnon peak
in La$_2$CuO$_4$. This reconfirms the magnitude of the ring-exchange
coupling $J_{\Box}=0.24J$ in this material. Ring-exchange creates additional
high energy spectral weight in the Raman intensity, but $J_{\Box}$ alone is
insufficient to explain the overall asymmetric lineshape and linewidth.

\begin{figure}[tbp]
\psfig{file=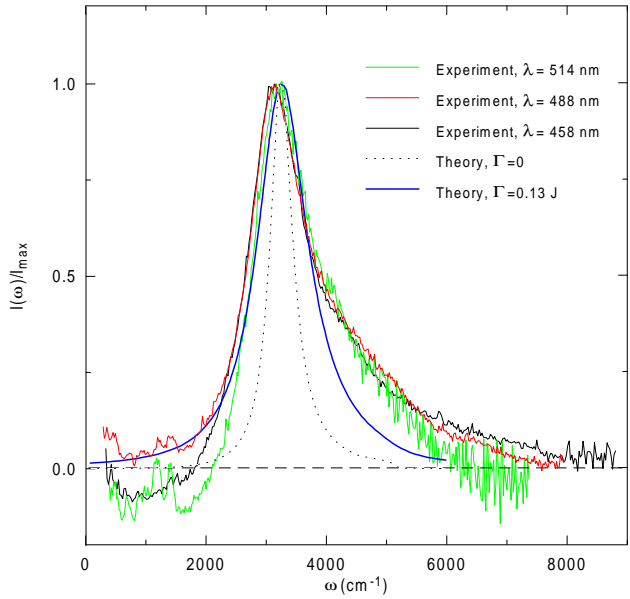,width=90mm,silent=}
%%\centerline{\psfig{file=Fig1.ps,width=90mm,silent=}}
\vspace{2mm}
\caption{Comparison of the theoretical result without ($\Gamma=0$) and with
additional damping ($\Gamma=0.13J$) to the experimental Raman intensity in La%
$_2$CuO$_4$ in B$_{1g}$ geometry taken from Ref. [3]. Calculations were
performed with the parameter set (16). The three experimental curves belong
to different incoming photon frequencies; the high-energy intensity
background was substracted. All data sets were normalized to their
corresponding intensity maximum.}
\label{fig:Ramanfit}
\end{figure}

It is a pleasure to thank T. Kopp and T. Nunner for insightful discussions.
We acknowledge support through Sonderforschungsbereich 484 of the Deutsche
Forschungsgemeinschaft.

\end{document}